# Group segmentation and heterogeneity in the choice of cooking fuels in post-earthquake Nepal

Ratna K. Shrestha[1,†,*] and Raunak Shrestha[2,†]

[1]Vancouver School of Economics, University of British Columbia, 6000 Iona Drive, Vancouver, BC, V6T 1L4, Canada

[2]Department of Radiation Oncology, University of California San Francisco, 1450 3rd Street, San Francisco, CA 94158, USA

[†]Equal contributions

[*]Correspondence should be addressed to R.K.S. (ratna.shrestha@ubc.ca)

**Corresponding author:**
Ratna K. Shrestha
Vancouver School of Economics,
University of British Columbia,
6000 Iona Drive, Vancouver, BC, Canada V6T 1L4
E-mail: ratna.shrestha@ubc.ca

## Abstract

**Segmenting population into subgroups with higher intergroup, but lower intragroup, heterogeneity can be useful in enhancing the effectiveness of many socio-economic policy interventions; yet it has received little attention in promoting clean cooking. Here, we use PERMANOVA, a distance-based multivariate analysis, to identify the factor that captures the highest intergroup heterogeneity in the choice of cooking fuels. Applying this approach to the post-earthquake data on 747,137 households from Nepal, we find that ethnicity explains 39.12% of variation in fuel choice, followed by income (26.30%), education (12.62%), and location (4.05%). This finding indicates that ethnicity, rather than income or other factors, as a basis of policy interventions may be more effective in promoting clean cooking. We also find that, among the ethnic groups in Nepal, the most marginalized Chepang/Thami community exhibits the lowest intragroup diversity (Shannon index = 0.101) while Newars the highest (0.667). This information on intra-ethnic diversity in fuel choice can have important policy implications for reducing ethnic gap in clean cooking.**

## Introduction

Widespread use of solid fuels, such as firewood, crop residue, and coal, for cooking is one of the major causes of health hazards[1–4], environmental degradation[5,6], and the gender gap[7,8] in many developing nations; it has also been linked to climate change [5,6,9]. The problem is so urgent that the United Nations has adopted universal access to clean fuels for cooking, such as liquefied petroleum gas (LPG), by 2030 as one of its sustainable development goals[10]. However, despite significant monetary and nonmonetary incentives for cleaner fuels and technologies, the rate of transition to clean cooking has failed to keep pace with population growth[11–13]; around 3 billion people worldwide, mostly from developing nations, still rely on solid fuels for cooking[14,15]. Against this backdrop, many studies[9,13,16,17] have recognized the importance of identifying target groups and designing policy instruments (such as promotional messages, subsidies to cleaner fuels, or nudges) to align with the fuel-choice behaviors of the groups. Nonetheless, the development of a systematic approach for group segmentation based on heterogeneity in households' fuel-choice decisions has received little or no attention. Here, we employ permutational analysis of variance (PERMANOVA), a distance-based multivariate analysis[18], for identifying the factor (such as ethnicity, income, or location) that segments households into subgroups with the highest intergroup dissimilarity in fuel-choice pattern.

Ad hoc or "one-size-fits-all" policy approaches may not be effective. For example, the Nepal government has been providing subsidies for biogas since 1990 to rural households but without targeting any specific ethnic or income groups. As a result, as of 2011, only about 5% of the households who were eligible for the subsidy adopted this fuel for cooking[19]. The progress report on clean cooking is not that encouraging in several other nations as well; solid fuels still constitute about 80% of cooking fuel in Sub-Saharan Africa[20], 76% in Northern Brazil[21], 74% in Guatemala[16], and 94% in rural Ghana[17]. To make matters worse, climate mitigation responsibilities, if fulfilled by imposing carbon taxes, can increase the costs of transition to cleaner fuels significantly[9]. To address this issue of slow transition to clean cooking, most of the studies[13,22–25] focus on regression analysis in determining the factors that significantly affect households' fuel choices. In this paper, we employ PERMANOVA, which is widely used in biological sciences[26–29], to determine the factor that captures the highest variation (given by sum of squared distances or, equivalently, $R^2$ values) in households' fuel-choice decisions.

Next, we measure fuel-choice diversity for each subgroup of the factor that captures the highest intergroup dissimilarity. For this purpose, we introduce Shannon index widely used in information theory and ecology. Shannon index measures the average number of binary questions (yes/no) that needs to be asked to determine the fuel choice of a randomly drawn household. Higher the value of the index, more diverse is the population subgroup[30] in their fuel-choice decisions. A zero diversity value of a subgroup would indicate that every household from that subgroup uses the same fuel type. The concept of diversity in energy or fuel choice, to the best of our knowledge, has not been studied before except in the context of energy security. In that context, it is considered as a proxy for energy security (or vulnerability) to unexpected supply disruption[30,31]. In our context, it can be considered as a proxy for unequal exposure to indoor pollution. Thus, the information on fuel-choice diversity may have important policy implications in closing the inequality in clean cooking and associated health outcomes across population subgroups.

The objective of this study therefore is twofold. First, it aims to determine the socio-economic factor that explains the highest intergroup dissimilarity in households' fuel-choice decisions. Second, it measures the fuel-choice diversity of each subgroup of the factor with the highest discriminatory power. To this end, we use data collected in the aftermath of Nepal's 7.8 Mw earthquake in 2015 by Kathmandu Living Labs for the government of Nepal.

## Results

### Types of cooking fuel used in earthquake-affected Nepal

Collected between January and March of 2016, the data we use for this study contain information on the types of energy use in cooking, impacts of the earthquake on buildings, and the socio-economic-demographic statistics of 747,137 households from 11 of the most-affected rural districts of Nepal (**Fig. 1A-B and Supplementary Fig. 1-3**). The data include individual households categorized into 96 social groups, 5 income classes, and 19 education levels. We regroup these 96 social groups into 10 major ethnic groups (**Fig. 1 and Supplementary Table 1**). Similarly, we regroup the education attainment of household heads into 5 major levels (**Fig. 1A and Supplementary Table 2**). Since the data include municipality (and ward) level location information, we divide the area into two geo-climatic regions: Himalayan and Hilly regions (**Supplementary Table 3**). The six types of fuel used in the survey area are: firewood, LPG, electricity, kerosene, gobar gas (biogas), and others (**Supplementary Table 4**).

Firewood is the dominant source of cooking fuel in all earthquake-affected area of rural Nepal. While over 87.83% of households use firewood as their primary source of cooking fuel, 10.88% use LPG, 1.18% biogas, 0.03% electricity, 0.03% kerosene, and 0.06% other fuels (**Fig. 1C and Supplementary Table 4**). Although households' fuel-choice decisions depend on multiple factors such as disposable income[22,23], ethnicity[25], cultural preferences[24], distance to forest[32], education[13,23], forest management regime[32,33], and gender of household head[13], among others, we investigate (due to data limitations) the effect of income, ethnicity, education of household head, and geo-climatic location. The data show that a large fraction of households from every ethnic group relies on firewood for cooking (**Fig. 1D**). In contrast, the prevalence of LPG, in general, increases with both income and education levels. Interestingly, however, among one of the ethnic groups, the Chepang/Thamis, the abundance of LPG use increases with income only up to Rs. 30,000-50,000 (US $1 = Nepali Rs. 110 approximately) per month; then it decreases

(**Fig. 1D**). The fuel-choice pattern by ethnicity and geo-climatic location and that by ethnicity and education level are displayed in **Supplementary Fig. 1B-C.**

**Intergroup dissimilarity and group segmentation**

To determine which of the four predetermined factors (ethnicity, income, education level of household head, or geo-climatic location) explains the most intergroup dissimilarity in fuel choice, we perform the Euclidean distance-based multivariate analysis. As mentioned before, ethnicity has ten levels, geo-climatic region has two levels, and the other two factors each has five levels. With the restructuring of the original data as outlined above, we obtain 4-factor ANOVA design that includes the number of households (abundance values) using each of the six fuel types by all combinations or treatments (**Supplementary Table 5**). Finally, we log transform the abundance data, compute the Euclidean dissimilarity matrix (based on pairwise differences in log-transformed abundance values between treatments), and perform PERMANOVA[18,34] on the dissimilarity matrix (see Methods).

Our findings show that intergroup dissimilarity in fuel choice is significant (Euclidean dissimilarity index, PERMANOVA, two-tailed significance level $\alpha = 0.05$, 999 permutations) across all four factors, with ethnicity explaining the highest intergroup dissimilarities ($R^2 = 0.391$, $P = 0.001$) followed by income ($R^2 = 0.263$, $P = 0.001$), education of household head ($R^2 = 0.126$, $P = 0.001$), and geo-climatic location ($R^2 = 0.041$, $P = 0.001$). To test the robustness of the above result, we also perform PERMANOVA based on Bray-Curtis dissimilarity matrix. Although the sum of squares and hence $R^2$ values are lower for all the four factors (than that based on the Euclidean dissimilarity matrix), their relative order of explanatory power remains the same. Thus, partitioning the households along ethnic lines for policy interventions may be more effective than along income class, geo-climatic location, or educational attainment.

However, PERMANOVA results can be confounded by the presence of within-group dispersions[35,36]. To investigate it for ethnic groups, we carry out a pairwise post-hoc test of the null hypothesis that there exists no differences in within-group dispersions. The Bonferroni corrected pairwise $P \geq 0.045$ for all pairs of ethnic groups show that there exists some within-group dispersions but such dispersion is not very high enough to invalidate the obtained PERMANOVA results (**Supplementary Table 6**).

Next, we perform ordination analysis to visualize the relative strengths of the four factors in explaining the variations in fuel-choice pattern. **Fig. 2A** shows the main effect of the four fixed factors represented by an unconstrained PCoA ordination of distances among the centroids of their respective levels. Consistent with the PERMANOVA results, the positions of the centroids of the ethnic groups (denoted by coloured circles) are more spread out along PCoA1 axis than the centroids representing income levels (diamonds), education levels (triangles), or geo-climatic locations (squares).

Subsequent to the above finding, we explore the PCoA ordination of the ethnic groups to gain further insight into the dissimilarity (or similarity) in fuel-choice behaviour. Interestingly, Nepal's most marginalized ethnic groups Chepang/Thamis (purple dots) are clustered in the upper-right quadrant, whereas historically most privileged groups such as Newars (red dots) and Brahmans (sky-blue dots) are mostly clustered on the lower-left quadrant (**Fig. 2B and Supplementary Fig. 4-5)**. **Fig. 2C** displays the positions of the centroids for ethnic groups (coloured circles) vis-à-vis the six fuel types (variables), which are represented by arrows. The lengths of the arrows are proportional to the correlation between the variables and the PCoA ordination. The above ordination results are corroborated by the heat-map cum dendrogram drawn by using Ward's minimum variance clustering method (**Fig. 2D**). This plot displays pairwise dissimilarity between the ethnic groups in the choice of all six fuel types. Notably, there exist two distinct clusters of ethnic groups. While Tamangs, Chettris, Brahmans, Newars, Gurung/Magars, and Dalits form one cluster, Rai/Limbus, Muslim/Others, Madhesis, and Chepang/Thamis form the other.

The finding that ethnicity has the highest discriminatory power may underscore the existence of an ethnic signature in fuel-choice decisions. But, it is unclear what underlying features of ethnicity, whether cultural practices, socio-cultural marginalization, or residency/tenancy situations, might have affected the variation in such behaviour. One of the ethnic groups, Madhesis, are recent migrants to the earthquake-affected area from Nepal's Terai region (the plains bordering India) and may not have access to firewood from community-managed forests as do indigenous communities such as Tamangs and Chepang/Thamis. Our results from post-earthquake Nepal are consistent with the findings from rural India[37] and Ethiopia[38], among others where household income is less significant in determining fuel choice compared to other social and cultural factors.

**Intragroup diversity and relative dependence**

Having identified ethnicity as the factor with the highest discriminatory power, we calculate intragroup (α) diversity for each ethnic group. **Fig. 3A** demonstrates that Chepang/Thamis exhibit the lowest α-diversity (median Shannon index = 0.101) followed by Tamangs, Dalits, and Rai/Limbus. Thus, most of the households from these communities rely on a single fuel (that is firewood). This indicates that the inequality in clean cooking, and hence associated health effects, is very high within these marginalized communities of Nepal. On the other hand, Newars exhibit the highest median Shannon index (0.667) followed by Brahman (0.656), indicating that these two historically privileged groups rely on diverse fuel types more equally than the other marginalized groups (**Supplementary Table 7**).

The Shannon indices of the individual ethnic groups are significantly different (Kruskal-Wallis test, $\chi^2 = 94.13$, df = 9, $P = 2.41 \times 10^{-16}$) (**Supplementary Table 8**). The pairwise comparison between Chepang/Thamis and Newars is the most significant (Mann-Whitney U test, Bonferroni corrected

P = $4.7 \times 10^{-11}$), whereas that between Tamangs and Dalits is the most nonsignificant (Wilcoxon-Mann-Whitney test, P = 0.94) (**Supplementary Table 8**).

However, α-diversity is a summary measure of intragroup diversity in that it can be lower also when an ethnic community excessively relies on any other type of fuel, instead of firewood. Thus while the information on intragroup diversity may have important policy implications in some context, relative dependence values (proportion of households using each fuel type) may be more relevant in others. For example, the relative dependence of an ethnic group on kerosene (transition fuel at the energy ladder[22,24] between solid fuels and LPG) may measure the group's likelihood to switch to LPG. Therefore, we calculate the relative dependence or abundance values for all ethnic groups (**Supplementary Table 9).** The highest proportion (33.24%) of Madhesis relies on LPG as their primary source of cooking fuel (**Fig. 3B)**. In contrast, the highest proportion (98.48%) of Chepang/Thamis relies on firewood.

Consistent with the results obtained above, Kruskal-Wallis tests show that ethnicity is significant in explaining the variation in the mean values of the relative dependence for all six fuel types (**Supplementary Table 10**). For example, variation in the dependence on firewood explained by ethnicity is significant (Kruskal-Wallis Test, $\chi^2 = 149.89$, df = 9, P = $9.30 \times 10^{-28}$). As for pairwise comparison (**Supplementary Table 11**), Chepang/Thamis exhibit the most differential relative dependence on firewood with all other ethnic groups, in particular with Newars (Wilcoxon-Mann-Whitney test, Bonferroni corrected pairwise P = $2.27 \times 10^{-9}$). The comparison between Dalits and Tamangs in firewood choice is statistically the most nonsignificant (Wilcoxon-Mann-Whitney test, P = 0.964), demonstrating that the mean values of the proportion of households using firewood from these two marginalized ethnic communities are not that different.

## Discussion

This study proposes a framework for identifying a factor that segments households into subgroups with the highest intergroup dissimilarity in the choice of cooking fuels. Employing this framework to the post-earthquake data from Nepal, we find that ethnicity explains the highest variation in fuel choice. This finding implies that partitioning households along ethnic lines (rather than along income class, educational attainment, or location) for policy interventions can be more effective.

While employing separate targeted policy specific to each ethnic subgroup is theoretically ideal, it could be prohibitively costly in practice. In contrast, a “one-size-fits-all” policy can be easier to implement, but likely to be ineffective in producing the desired results. So, an appropriate middle ground could be to bundle subgroups into a manageable number of larger target group(s). Such bundling can be based on the pairwise similarity (or dissimilarity) in the choice of all six fuel types (**Fig. 2D**) or any single fuel type (**Fig. 3B**). Alternatively, the Shannon index values (**Fig 3B**) may be used in regrouping the subgroups into one combined target group. For example, given a particular promotional message as a policy tool, bundling two ethnic communities Dalits and Tamangs (with similar Shannon index values) into one target group can be more effective than bundling Chepang/Thamis and Newars (with entirely two different Shannon values). Shannon index values of ethnic groups may also have important policy implications for closing the intra-ethnic disparity in clean cooking and the resulting unequal health outcomes (such as respiratory illness and lung cancer)[8,39].

Our results may also be helpful in determining the kind of policy interventions (such as awareness campaigns or behavioral incentives such as nudges) that the target group is more likely to respond.[9,16] Just because a certain household subgroup overwhelmingly relies on firewood (or kerosene for that matter) does not mean that the subgroup is likely to respond to a particular policy intervention; the subgroup’s response might be guided by cultural practices, personal tastes, and cooking habits[16,24]. For example, Chepang/Thamis, by virtue of being indigenous[40] to the survey area, may respond to the awareness campaigns on the importance of forest conservation more than similar campaigns on the harmful health effects of cooking with firewood.

Whereas the focus of this study is to identify a single criterion of segmentation, the conceptual framework presented can be readily extended to encompass multiple criteria. In fact, policies based on multiple criteria, although impractical, can theoretically produce better results. However, group heterogeneity can be sensitive to the basis chosen for classifying households (e.g., religion instead of ethnicity, rural/urban instead of geo-climatic location, etc.) ex ante, and not to mention the number of levels chosen (e.g., number of income, ethnic, or education levels). Therefore, the question of which criterion (or criteria) of group segmentation yields the highest intergroup heterogeneity is purely empirical.

## Methods

**Dataset:** Following the 7.8 $M_w$ Gorkha Earthquake in Nepal on April 25, 2015, the Kathmandu Living Labs in collaboration with the National Planning Commission (the Government of Nepal), carried out the largest household survey ever done in Nepal using mobile technology. Although the primary objective of this survey was to assess damages inflicted by the quake and identify beneficiaries eligible for government's housing reconstruction grants, the data contain many other kinds of valuable socio-economic information, including the types of fuel used by households for cooking and lighting from 11 of the most earthquake-affected districts of Nepal, excluding the Kathmandu valley. The survey was carried out between January and March of 2016, about 10 months after the quake that killed over 9,000 people.

The data for all 11 districts were downloaded from the 2015 Nepal Earthquake: Open Data Portal (URL: http://eq2015.npc.gov.np/) in September 2018. For this study, we use the following two data files: (i) csv_household_demographics.csv; this file contains information on household size, ethnicity, household income, and gender, education and age of household head, and (ii) csv_household_resources.csv; this file contains data on the type of fuel used by households for cooking and lighting, source of drinking water, and the ownership of electronic appliances and motor vehicles. These files include the information on 747,365 households. Since some data on 228 households were missing, we removed them and carried out our analysis with the remaining 747,137 observations.

**Data Preparation:** We restrict our analysis to four key socio-economic and demographic information of the households (ethnicity, income, education, and location) and their post-earthquake sources of cooking fuel. The dataset includes the types of cooking fuel used by 96 different social groups found in the area. Since the relative population sizes of many of these groups are small, we regroup them together to form a larger group with comparable population size; we also merge some of them with the larger major groups to which they are closer culturally or linguistically[40]. With this regrouping, we obtain 10 major ethnic groups: Brahman, Chepang-Thami, Chettri, Dalit, Gurung-Magar, Madhesi, Newar, Rai-Limbu, Tamang, and Muslim-Others (**Supplementary Table 1**).

The education level of the household heads range from illiterate to Doctoral degree. We regroup the education attainment into 5 levels: Illiterate; Non-formal and Others; grade 1-7 as Primary; grade 8-12 as Secondary; and Bachelors, Masters, and PhDs as University level education (**Supplementary Table 2**). Similarly, the data provide information on municipality (and ward-level) location which allows us to divide the entire earthquake-affected 11 contiguous districts into two geo-climatic regions: Himalayan and Hilly regions. Approximately Northern half of the affected areas dotted mostly with mountains that are higher than 3,000 m in elevation are classified as Himalayan region and the rest of the Southern part as Hilly region (**Supplementary Table 3**)[41]. As for income levels, we follow the original dataset's categorization that breaks the monthly income of the households into 5 levels: Nepali Rs. 10,000 or less; Rs. 10,000 to 20,000; Rs.

20,000 to 30,000; Rs. 30,000 to 50,000; and more than Rs. 50,000 (US $ 1 = about Nepali Rs.110). With the way we have regrouped the factors (ethnic classes or education levels), we do not mean to offer any particular process of regrouping. Rather, given multiple factors and levels considered for policy intervention ex ante, we offer a framework for identifying a single factor of segmentation that enjoys the highest discriminatory power.

With this restructuring of the data, we obtain an ANOVA factorial design, with 4 fixed factors: ethnicity, income, geo-climatic region, and education of household head. As explained above, ethnicity has 10 levels, geo-climatic region has 2 levels while income and education has 5 levels each. Thus, by design, our ANOVA table is supposed to have 10x2x5x5 = 500 different combinations or treatments. However, in one of the ethnic groups, the Chepang-Thami, there are no households with university education and income more than Rs. 50,000 per month. Also in this ethnic group, no household with non-formal/other education and income higher than Rs. 50,000 per month reside in Hilly region. Therefore, our ANOVA table consists of 497, instead of 500, combinations or treatments (**Supplementary Table 5**).

**Statistical Analyses:** All analyses are performed using R (version 3.5.1) and statistical functions embedded in the Vegan R-package[34,42].

*Intergroup dissimilarity*: Dissimilarity between any two treatments is assessed with Euclidean dissimilarity matrix (Vegdist function in the Vegan R-package) on log-transformed abundance data (the number of households using each of the 6 fuel types); this matrix gives pairwise Euclidean distances in fuel-choice behaviour among all 497 treatments. Euclidean distance between treatment j and k is calculated as below.

(1) $$E_{jk} = \sqrt{\sum_{i=1}^{n}(X_{ij} - X_{ik})^2}$$

where, n = total number of fuel types, $X_{ij}$ = log of number of households using fuel type i in treatment j, and $X_{ik}$ = log of number of households using fuel type i in treatment k. The log transformation is used to reduce the effect of excessively high proportion of certain ethnic groups (such as Tamangs) and income levels (less than Rs. 10,000 per month) in the population. We use a Euclidean distance because it is neutral to the presence of double zeros (i.e. does not affect dissimilarity values). Unlike in ecology where double zeros (i.e. no presence of species) can occur when two sites have entirely different situations, for instance climates that are too hot or too cold, such issues are nonexistent in the present analysis. Moreover, Euclidean distance preserves the original dissimilarities and poses no issue with negative eigenvalues[43].

The contribution of each factor (ethnicity, income class, geo-climatic location, and education) on the dissimilarity matrix (dependent variable) is calculated by performing PERMANOVA[18,34,44]. We use adonis2() function with 999 permutations[34,42]. The sum of squares values (or equivalently $R^2$) measure the extent of dissimilarity explained by each factor[18,44]. We use PERMANOVA (non-parametric method) because our data do not satisfy normality assumption. We report only the additive effects of the aforementioned 4 factors and not the interaction effects among them (as the interaction effect between ethnicity and income explains only about 2.76% of dissimilarity in the data). The null hypothesis of this test $H_0$ is: the locations of centroids of the factors do not differ (**Supplementary Table 6**). We also repeat this analysis with Bray-Curtis dissimilarity matrix to see if we obtain similar results as with Euclidean matrix. The Bray-Curtis dissimilarity between two treatment units is given by

(2) $$BC_{jk} = \frac{\sum_{i=1}^{n} |x_{ij} - x_{ik}|}{\sum_{i=1}^{n} (x_{ij} + x_{ik})}$$

The implicit assumption behind PERMANOVA is that there exists no within-group dispersion. To test whether the PERMANOVA results are artifacts of within-group dispersions[35,44], we perform post-hoc pairwise tests (of the null hypothesis that there exists no differences in within-group dispersions among all pairs of ethnic groups) by using pairwise.perm.manova function of RVAideMemoire package for R with Bonferroni multiple-testing corrections[45] (**Supplementary Table 7**).

We also validate our PERMANOVA results by performing Principal Coordinate Analysis (PCoA) ordination (Vegan R-package) on the Euclidean dissimilarity matrix. The 6 fuel types shown by arrows (**Fig. 2c**) are drawn using Vegan: envfit. The length of arrow of each fuel type measures its relative contribution to the dissimilarity in fuel-choice pattern. Heat-map cum dendrogram for unsupervised hierarchical clustering (using Ward's minimum variance method on Euclidean distance) is generated with gplots (R package).

*α-diversity:* Shannon diversity indices are calculated on raw abundance data (Vegan R-package). Shannon Index for a subgroup $k$ ($k$ = 1, .., $K$), $H_k$, is given by the following equation:

(3) $$H_k = -\sum_{g=1}^{K} p_{gk}\, ln(p_{gk})$$

where $g$ is fuel type ($g$ = 1, …K) and $p_{gk}$ is the proportion of households using $g^{th}$ fuel in subgroup k and is equal to $n_{gk}/n_k$, with $n_{gk}$= number of households from subgroup k using $g^{th}$ fuel and $n_k$ = total number of households in subgroup k. Shannon index measures both richness (number of fuel types) and evenness (the distribution of the fuel types).

To test whether the diversity between ethnic groups is significant ($H_0$: the difference in the median values of Shannon index of all the ethnic communities are equal to zero), we perform Kruskal-Wallis test by ranks (1 test for each of the 6 fuel types), a nonparametric method, as our data do not follow normality assumption.

Kruskal-Wallis tests, however, do not identify which pairs of ethnicities have significantly different median relative abundance values. So, we also carry out pairwise comparison (post-hoc tests) using Wilcoxon-Mann-Whitney test with Bonferroni correction.

*Relative abundance values*: We use total transformation (Vegan R-package) on the raw abundance values (number of households using different fuel types) to obtain relative abundance values. Due to excessively high difference in population between certain ethnic groups in the survey area (see **Supplementary Table 9**), it was logical to analyze with relative abundance values instead of the raw values.

To test the null hypothesis ($H_0$ = differences in the mean relative abundance values are zero), we perform Kruskal-Wallis tests. We also perform post-hoc pairwise comparison using Wilcoxon-Mann-Whitney tests with Bonferroni correction.

*Choropleth map*: We obtained the geographical coordinates of Nepal's administrative units/districts in GeoJSON format from the Open Knowledge Nepal data portal http://localboundries.oknp.org/ (accessed in February 2019). We parsed the GeoJSON data using "geojsonio" R-package and generated the choropleth maps with "ggplot2" R-package.

## Acknowledgements

We are grateful to Kathmandu Living Labs and the Government of Nepal for making the survey data publically accessible. We thank Patrick Francois (Economics) and Sara Shneiderman (Anthropology), from University of British Columbia and Robert Bell from Vancouver Prostate Centre for helpful comments. We also thank Anish Joshi and his team from Genesis Consultancy Pvt. Ltd. Kathmandu, Nepal for their help in producing Choropleth maps.

## Data availability

The datasets used in this study are available at the 2015 Nepal Earthquake: Open Data Portal (http://eq2015.npc.gov.np/)

## Code availability

R-codes used for data analysis is available at https://raunakms.github.io/diversity_cooking_fuel and can be freely downloaded from https://github.com/raunakms/diversity_cooking_fuel

## Author contributions

Both authors conceived the study, analyzed the data, and wrote the manuscript.

## Competing interests

The authors declare no competing interests.

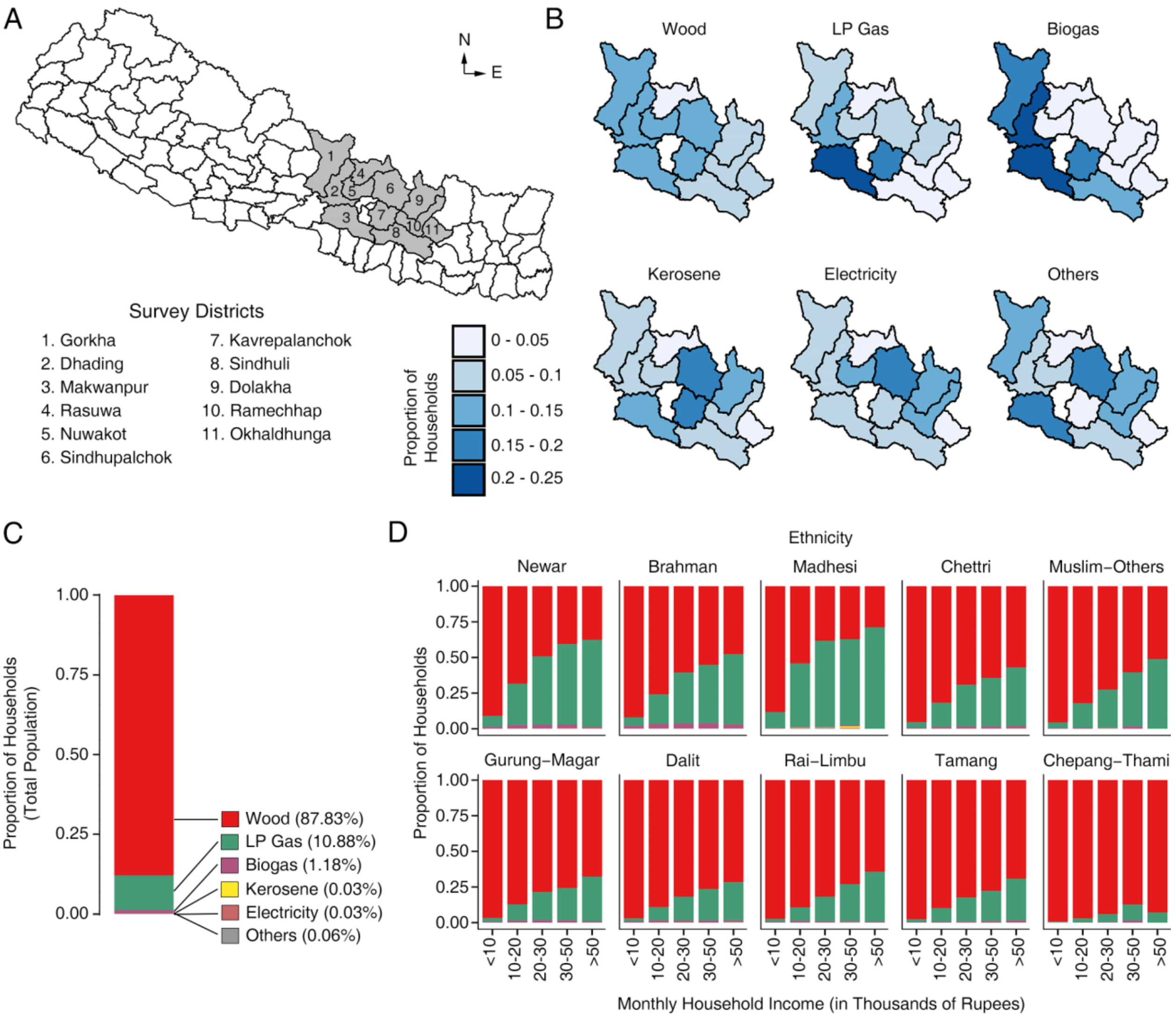

**Fig. 1: Distribution of fuel-types across earthquake-affected districts of Nepal. A)** The map of Nepal and the locations of 11 earthquake-affected districts. The middle 3 districts (shown as blanks) are Kathmandu, Lalitpur, and Bhaktapur, for which no data are available. **B)** Choropleth map showing the relative abundance of 6 fuel-type across the 11 districts. **C)** Proportion of the total population using different cooking fuels. **D)** Proportion of the households using different cooking fuels segregated by ethnicity and household income.

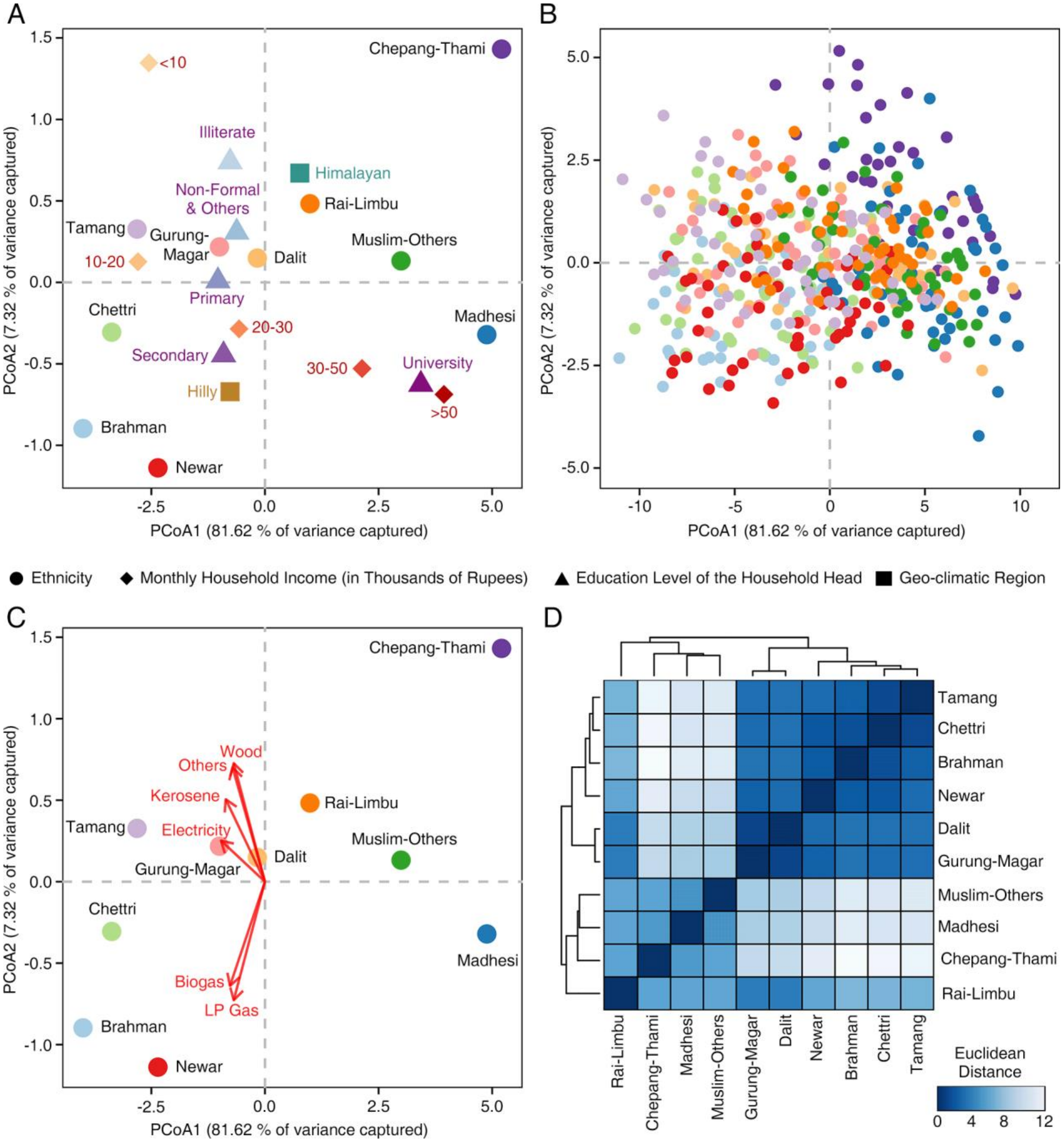

**Fig. 2: PCoA Ordination based on a Euclidian distance of log-transformed data. A)** PCoA ordination showing the centroids of individual levels of each of the four fixed factors: ethnicity, income, education of household head and geo-climatic region. Segmenting the households by ethnicity displays the highest dissimilarity (PERMANOVA, $R^2$ = 0.39) followed by income ($R^2$ = 0.26), education ($R^2$ = 0.12), and geo-climatic location ($R^2$ = 0.04). The positions of the circles

(representing the centroid of each ethnicity) are more spread out along PCoA1 than the positions of the diamonds (income levels), the triangles (education levels), and the squares (geo-climatic regions). **B)** PCoA ordination displaying the dissimilarity in fuel-type choice by ethnicity only. **C)** PCoA biplot (Vegan R package: envfit) with colored circles representing centroids of ethnic groups and arrows representing six types of fuel that contribute to the dissimilarity. **D)** Heat-map displaying the pairwise Euclidean dissimilarity among the ethnic groups.

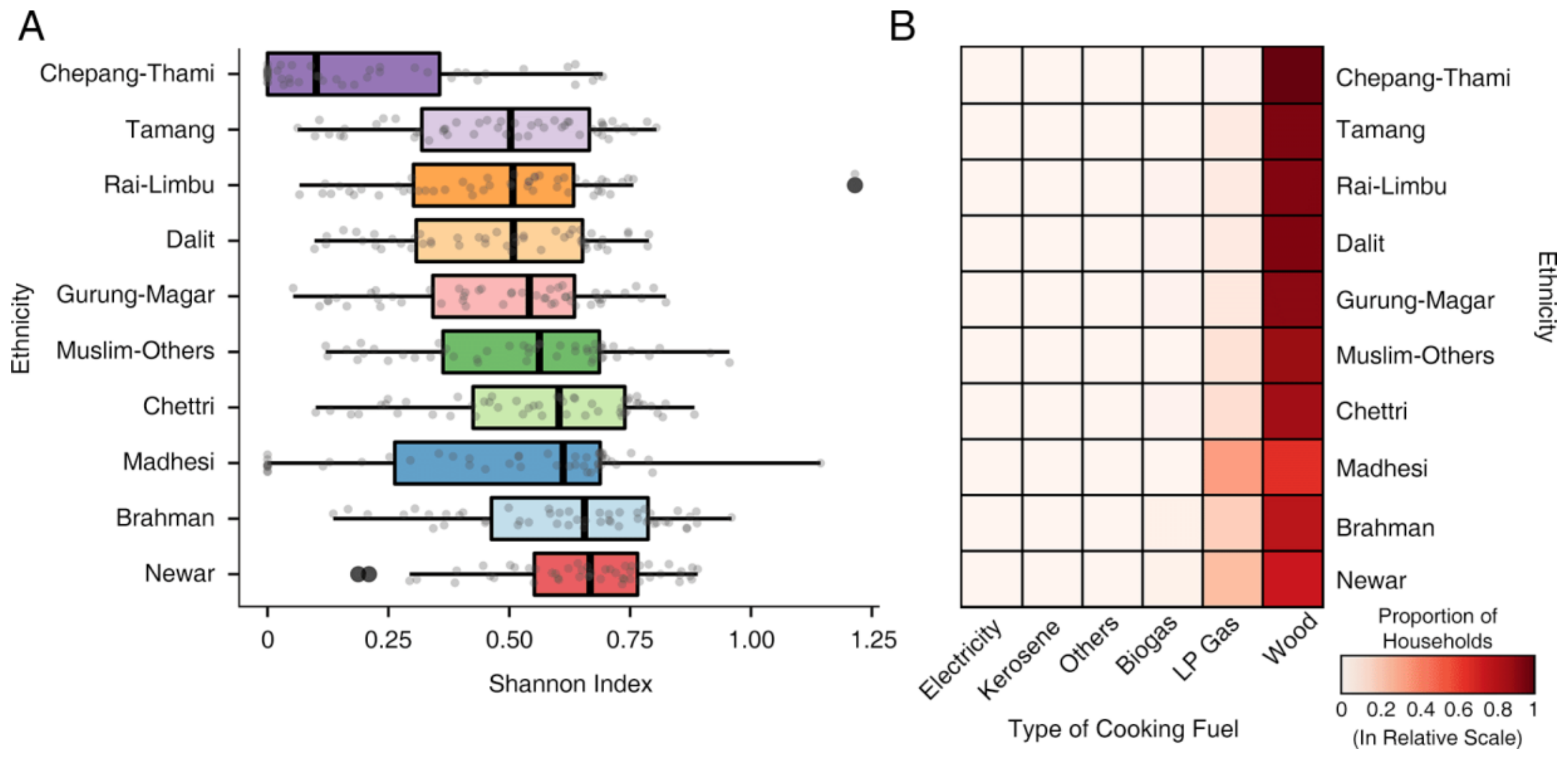

**Fig. 3: Shannon indices and relative abundances by ethnic groups on raw data**. **A)** The boxplot of the Shannon indices of each ethnic group. The boxplot displays the median (center line) with box limits 75% upper and 25% lower quartiles. The darker points outside the whiskers, denoting 1.5x interquartile range, represent outliers. The grey dots show the relative positions of individual treatments. **B)** Heat-map demonstrating the proportion of households from each ethnic group using each of the 6 different fuel types.

**Table 1.** Summary of PERMANOVA analysis comparing the intergroup dissimilarity in fuel-choice pattern. Both Euclidean and Bray-Curtis distance-based analyses show that ethnicity has the highest discriminatory power in fuel-choice pattern.

| | **Euclidean Distance** | | | **Bray-Curtis Distance** | | |
|---|---|---|---|---|---|---|
| | **F** | **$R^2$** | **p-value** | **F** | **$R^2$** | **p-value** |
| Ethnicity | 115.805 | 0.391 | 0.001 | 46.34 | 0.329 | 0.001 |
| Income | 175.183 | 0.263 | 0.001 | 59.76 | 0.187 | 0.001 |
| Education | 84.030 | 0.126 | 0.001 | 28.13 | 0.008 | 0.001 |
| Geo Region | 107.978 | 0.041 | 0.001 | 27.44 | 0.002 | 0.001 |
| Residual | - | 0.179 | - | - | 0.374 | - |

# Supplementary Figures

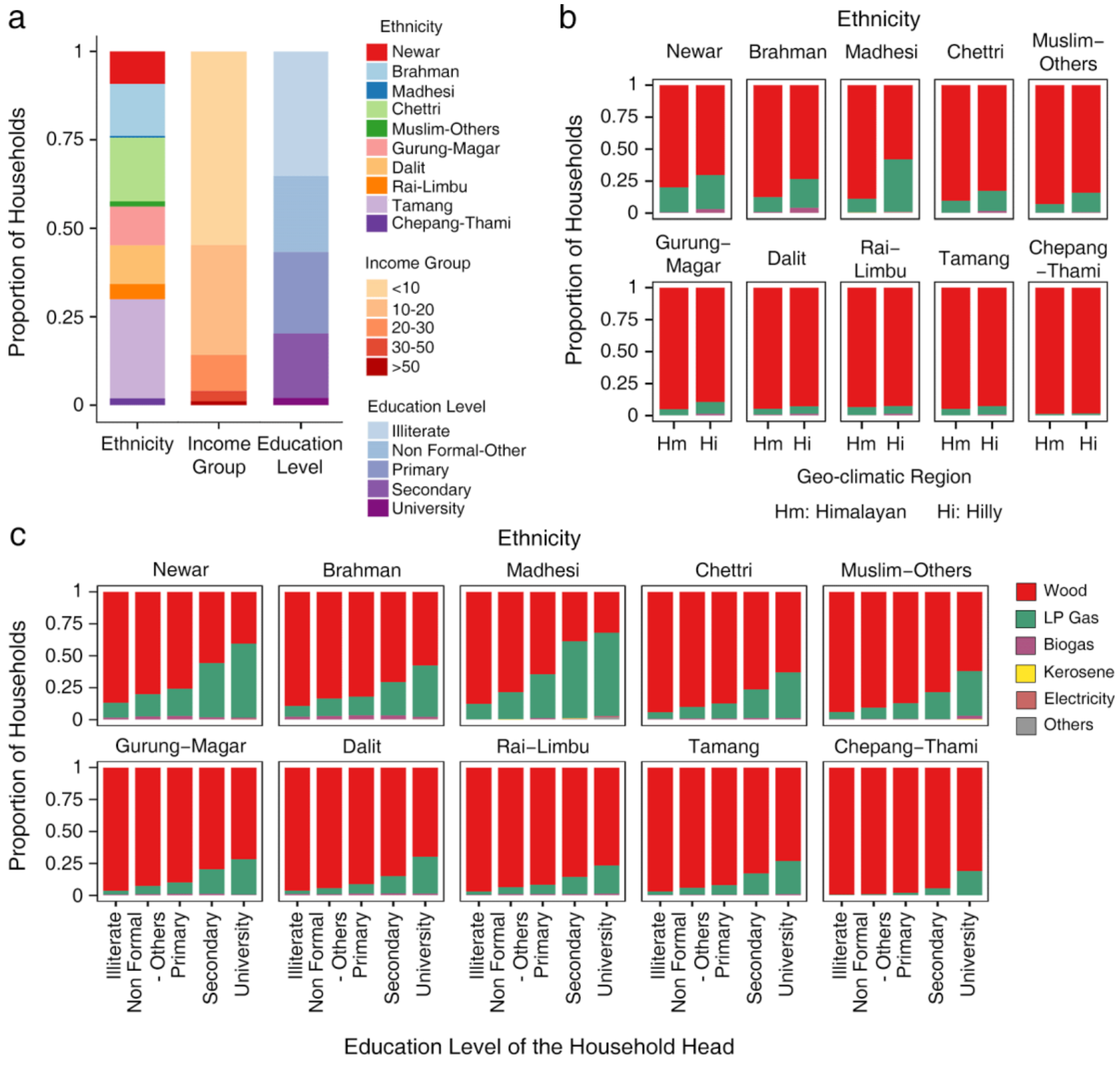

**Supplementary Fig. 1: Proportion of households in the study dataset. a)** Proportion of households by ethnicity, monthly household income, and education level of the household head. **b)** Proportion of households using different types of cooking fuel grouped by ethnicity and geo-climatic region, and **c)** Proportion of households using different types of cooking fuel grouped by ethnicity and education level of the household head.

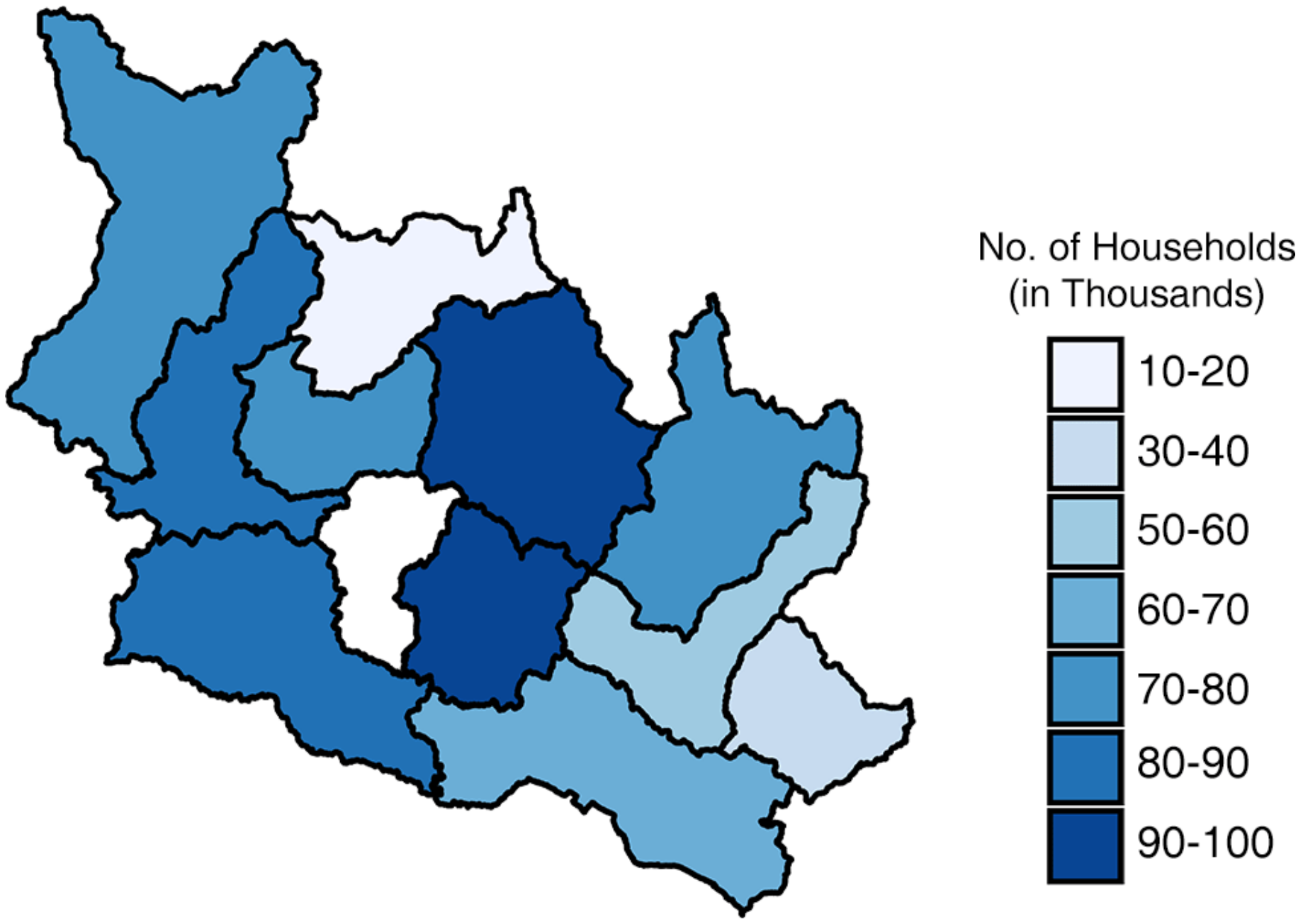

**Supplementary Fig. 2: Choropleth map showing the distribution of households in the 11 of the earthquake-affected rural districts of Nepal.** The shades of color represent the total number of households in the respective district. The blank space at the centre is Kathmandu Valley for which no data are available.

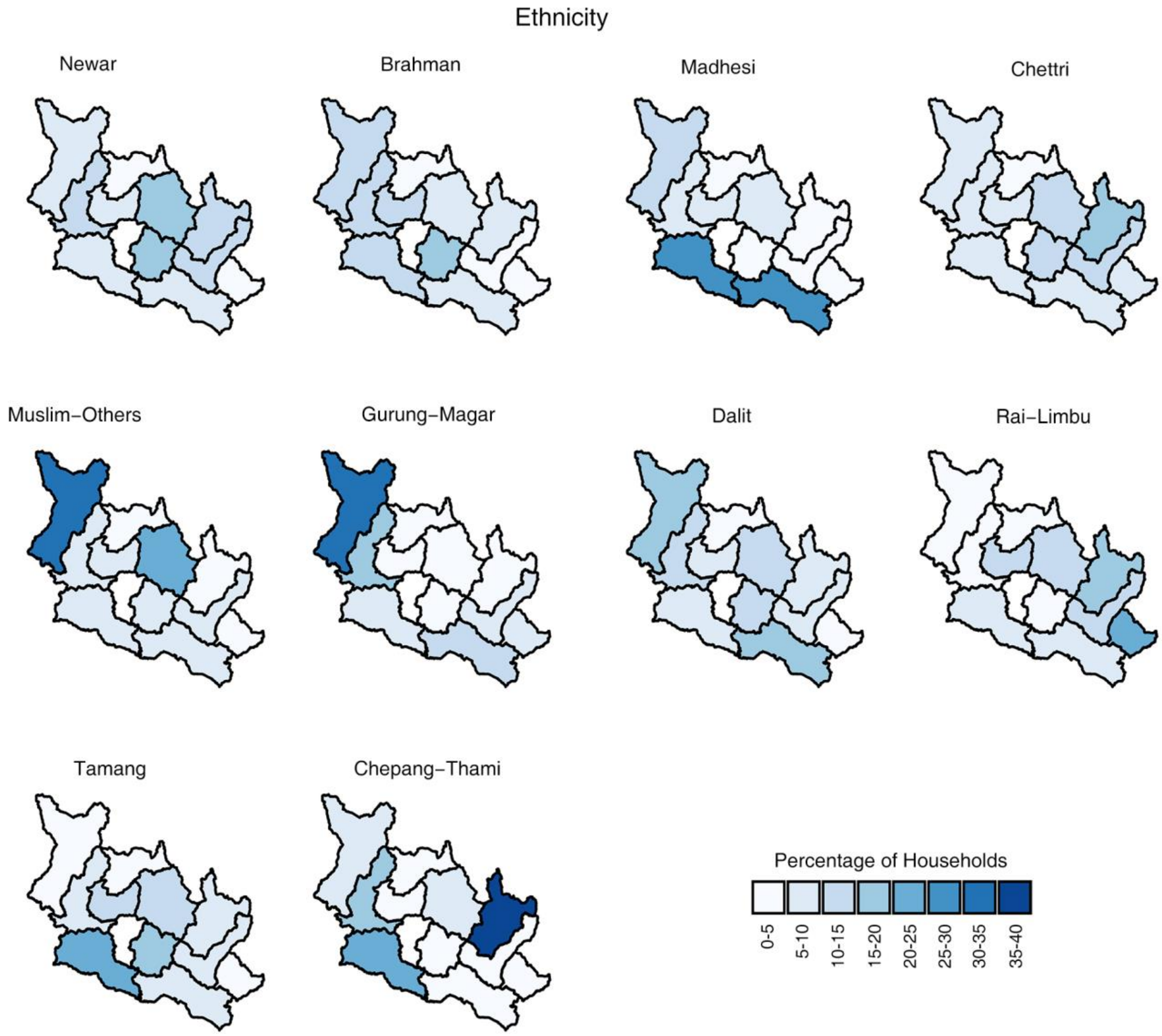

**Supplementary Fig. 3: Choropleth map showing the distribution of households by ethnicity in the 11 of the earthquake-affected rural-districts of Nepal.** The shades of color represent the percentage of households. For example, about 35-40% of Chepang-Thamis live in Dolakha district (darkest colour).

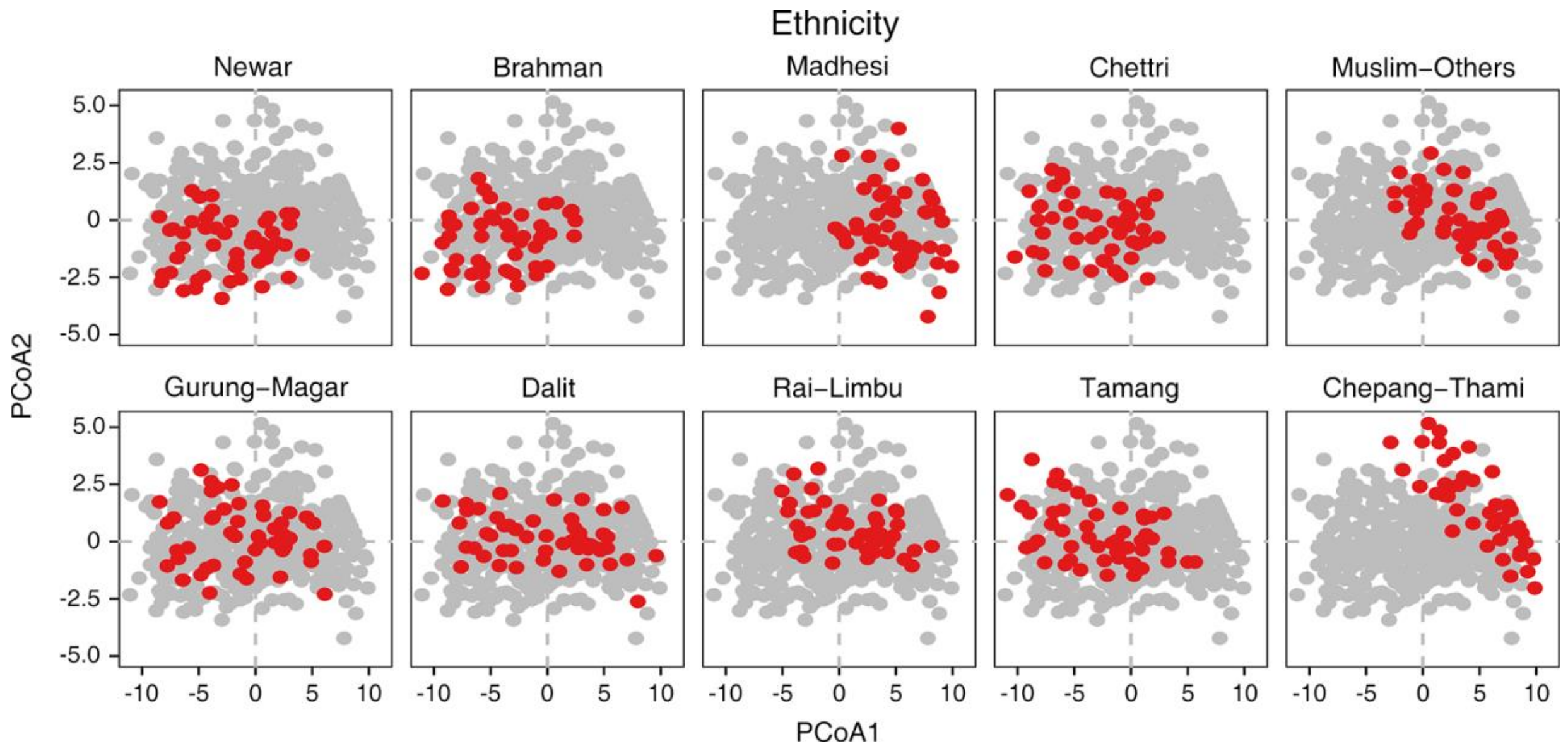

**Supplementary Fig. 4: PCoA ordination showing the positions of individual treatments of ethnic groups**. Each ethnic group has 50 individual treatments, except Chepang-Thami which has 47.

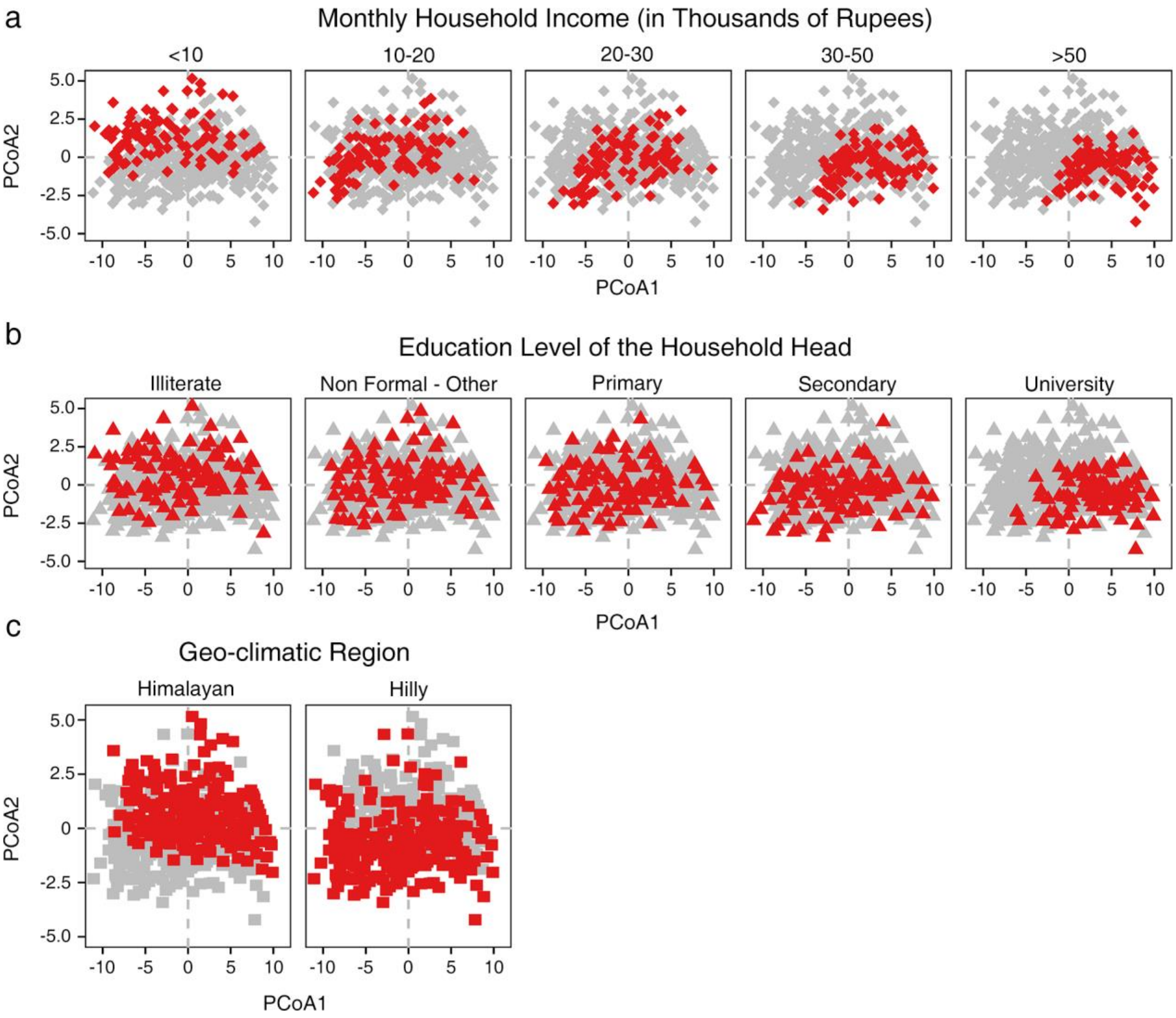

**Supplementary Fig. 5: PCoA ordination displaying the positions of each individual treatment of the 3 factors**. **a)** income groups, **b)** education levels of household head, and **c)** geo-climatic regions.